\documentclass[%
 aip,
 amsmath,amssymb,
 reprint,%
]{revtex4-1}

\usepackage{graphicx}
\usepackage{dcolumn}
\usepackage{bm}
\usepackage{xcolor}
\usepackage[utf8]{inputenc}
\usepackage[T1]{fontenc}
\usepackage{mathptmx}
\usepackage{etoolbox}

\makeatletter
\def\@email#1#2{%
 \endgroup
 \patchcmd{\titleblock@produce}
  {\frontmatter@RRAPformat}
  {\frontmatter@RRAPformat{\produce@RRAP{*#1\href{mailto:#2}{#2}}}\frontmatter@RRAPformat}
  {}{}
}%
\makeatother
\begin{document}

\preprint{AIP/123-QED}

\title[Engineering multiple GHz mechanical modes in optomechanical crystal cavities]{Engineering multiple GHz mechanical modes in  optomechanical crystal cavities}
\author{Laura Mercad\'e}
\affiliation{Nanophotonics Technology Center, Universitat Polit\`ecnica de Val\`encia, Camino de Vera s/n, 46022 Valencia, Spain}
\affiliation{MIND-IN2UB, Departament d’Enginyeria Electrònica i Biomèdica, Facultat de Física, Universitat de Barcelona, Martí i Franquès 1, Barcelona 08028, Spain}
\author{Raúl Ortiz}%
 \affiliation{Nanophotonics Technology Center, Universitat Polit\`ecnica de Val\`encia, Camino de Vera s/n, 46022 Valencia, Spain}
\author{Alberto Grau}
\affiliation{Nanophotonics Technology Center, Universitat Polit\`ecnica de Val\`encia, Camino de Vera s/n, 46022 Valencia, Spain}
\author{Amadeu Griol}
\affiliation{Nanophotonics Technology Center, Universitat Polit\`ecnica de Val\`encia, Camino de Vera s/n, 46022 Valencia, Spain}
\author{Daniel Navarro-Urrios}
\affiliation{MIND-IN2UB, Departament d’Enginyeria Electrònica i Biomèdica, Facultat de Física, Universitat de Barcelona, Martí i Franquès 1, Barcelona 08028, Spain}
\author{Alejandro Mart\'inez}
\affiliation{Nanophotonics Technology Center, Universitat Polit\`ecnica de Val\`encia, Camino de Vera s/n, 46022 Valencia, Spain}
\email{amartinez@ntc.upv.es}

\date{\today}

\begin{abstract}
 Optomechanical crystal cavities (OMCCs) are fundamental nanostructures for a wide range of phenomena and applications. Usually,  optomechanical interaction in such OMCCs is limited to a single optical mode and a unique mechanical mode. In this sense, eliminating the single mode constraint - for instance, by adding more mechanical modes - should enable more complex physical phenomena, giving rise to a context of multimode optomechanical interaction. However, a general method to produce in a controlled way multiple mechanical modes with large coupling rates in OMCCs is still missing. In this work, we present a route to confine multiple GHz mechanical modes coupled to the same optical field with similar optomechanical coupling rates - up to 600 kHz - by OMCC engineering. In essence, we increase the number of unit cells (consisting of a silicon nanobrick perforated by a circular holes with corrugations at its both sides) in the adiabatic transition between the cavity center and the mirror region. Remarkably, the mechanical modes in our cavities are located within a full phononic bandgap, which is a key requirement to achieve ultra high mechanical Q factors at cryogenic temperatures. The multimode bevavior in a full phononic bandgap and the easiness of realization using standard silicon nanotechnology make our OMCCs highly appealing for applications in the classical and quantum realms.
\end{abstract}

\maketitle


\section{Introduction}

Cavity optomechanics, the scientific field that studies the interaction between light and mechanics in solid cavities \cite{ASP14-RMP,PEN14-NP}, usually considers the coupling of a single mechanical mode with a single optical field \cite{CHAN11-NAT,GRO09-NAT,VER12-NAT}. But more complex physics and novel phenomena may arise when considering multiple mechanical modes coupled to a single optical mode \cite{SCH17-PNAS,PINO22-NAT,MER21-PRL}. Among others, the interest in multimode self-oscillating systems has led to emergent phenomena including dynamical topological phases \cite{WAL16-NJP}, analog simulators \cite{MAH16-SA}, synchronization \cite{HEI11-PRL,ZHA12-PRL,HOL12-PRE,LOR17-PRL,COL19-PRL,PEL20-PRR,MA20-ARX} and stability enhancement \cite{ZHA15-PRL}, with the last two applications employing multiple confined mechanical modes coupled through an optical field.

In many of these multimode systems, the involved mechanical modes are not confined into the same physical structure. This is the case, for example, of confined mechanical modes in different optomechanical crystal cavities (OMCCs) that are coupled via mechanical interaction \cite{COL19-PRL} or coupled micromechanical oscillators that interact through an optical radiation field but are also physically localized at different resonators \cite{PRL17-GIL, ZHA12-PRL, ZHA15-PRL}. Conversely, multiple mechanical modes confined in the same structure and interacting with one common intracavity optical field have also been studied in the literature for a given number of mechanical oscillators \cite{PRL20-SHE, MAT20-NN, LEI17-NCOMM}. Remarkably, most of these systems involve oscillators up to MHz frequencies, so a general route towards multiple GHz mechanical resonances in a single cavity is still missing. Recently, a bullseye optomechanical (OM) resonator enabling multiple mechanical and optical modes has been fabricated and tested \cite{SAN17-OE}, but the vacuum OM coupling rates are about one of magnitude smaller than in silicon OMCCs \cite{CHAN11-NAT,MER20-NN}.

For some applications, it would also be interesting to have all these mechanical modes placed in a full phononic bandgap that prevents phonons to escape from the cavity. This could be particularly helpful in quantum applications at cryogenic temperatures, because of the enhancement of the mechanical Q factor in these conditions \cite{MAC19-ARX}. 
A reliable route to multiple phonon modes with large coupling rates would also open the door to versatile all-optical OM-based microwave signal synthesis \cite{MER20-NN} and processing \cite{MER21-LPR}, which is especially valuable for application in wireless systems, in particular those requiring extreme compactness and lightweight (satellite communications) \cite{MER22}.

Here, we propose a general method to get multiple mechanical modes with GHz frequencies and placed in a full phononic bandgap of a silicon OMCC. Starting with the design in \cite{OUD14-PRB} and demonstrated experimentally in 
\cite{MER20-NN}, we go a step further and demonstrate that engineering the central region of the cavity enables to systematically increase the number of mechanical modes whilst ensuring relatively large values of the OM coupling rate ($g_0$/2$\pi$ $\approx 400 kHz$). Although we use silicon as underlying material, this approach should also be of application when using other high refractive index materials to build the cavities. Such large values of $g_0$ allows us to easily transduce all the confined modes into the driving optical field. We demonstrate experimentally the existence of up to six modes in a single cavity, though our method could be used to get even more mechanical modes. All these features, together with the fabrication on a silicon chip using standard nanofabrication tools, suggest that our cavities could play a role in the development of multimode cavity optomechanics for different classical and quantum applications.


\section{Photonic and phononic band diagrams of the OMCC}
\begin{figure}
    \centering
    \includegraphics[width=\columnwidth]{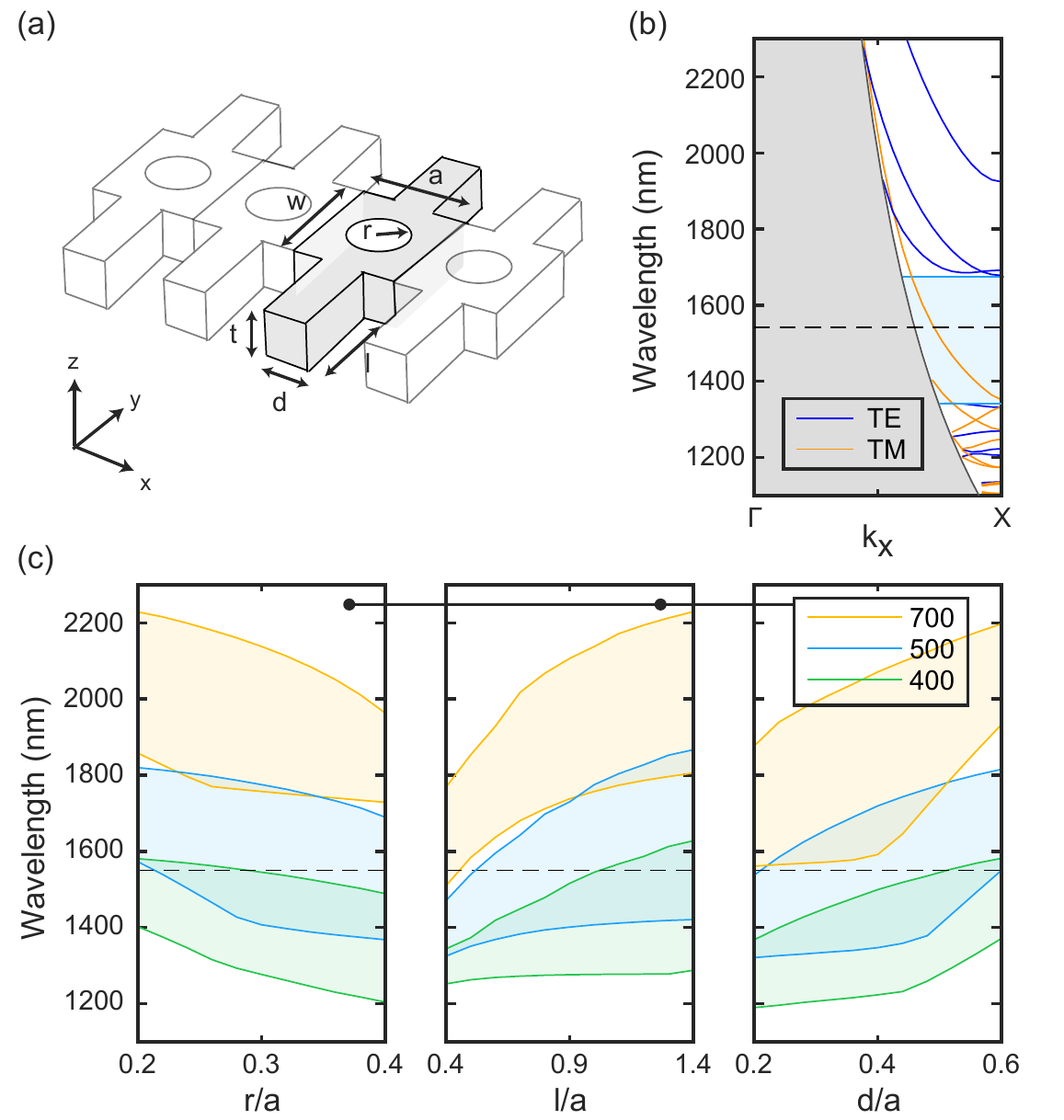}
    \caption{Photonic band calculations of the mirror unit cell. (a) Array of the unit cells under study depicting the main parameters. (b) Photonic band diagram of the mirror unit cell for a lattice period $a$=500 nm, $t$=220 nm, $l/a=$0.9, $d/a=$0.4 and $r/a=0.3$. (c) Evolution of the total photonic bandgap of different aspect ratios for three different lattice periods (shown in the inset, values in nm). From left to right, the results shown in the panels are obtained for the ratios $l/a=$0.9 and $d/a=$0.4 (first panel), $r/a=$0.3 and $d/a=$0.4 (second panel) and $r/a=$0.3 and $l/a=$0.9 (third panel), where $d$, $r$ and $l$ are geometrical parameters of the cavity cell showed in Fig. \ref{fig:photobands}(a).}
    \label{fig:photobands}
\end{figure}

Prior to the design of the defect cells and the cavity modes, we perform a broad analysis of the mirror region of the OMCC to optimize both the photonic and phononic bandgaps. Figure \ref{fig:photobands}(a) sketches the unit cell used to build up our cavity. It consists of a 220 nm thick silicon nanobrick drilled with a circular hole and surrounded by lateral stabs or corrugations. The resulting photonic band diagram of this unit cell is depicted in Fig. \ref{fig:photobands}(b), showing the TE-like (in blue) and the TM-like (in orange) bands, respectively. The gray shadowed area corresponds to the light cone and the blue shadowed area, between the third and fourth TE-like bands, denotes the quasi-TE bandgap, which will be used to confine the driving optical mode. It must be noted that this unit cell provides a large tunability in wavelength, as can be seen in Fig. \ref{fig:photobands}(c), which depicts the evolution of the first TE bandgap as a function of the aspect ratio of the unit cell for three different period lattices. A broad tunability for wavelengths between 1200 nm and 2200 nm can be obtained, which provides a great versatility in the design.  As our objective is to confine an optical mode around 1550 nm (depicted as a dashed line in Fig. \ref{fig:photobands}(b,c)), from now on we will focus our studies in a period lattice for the mirror part of the cavity with $a=$500 nm.  

\begin{figure}
    \centering
    \includegraphics[width=\columnwidth]{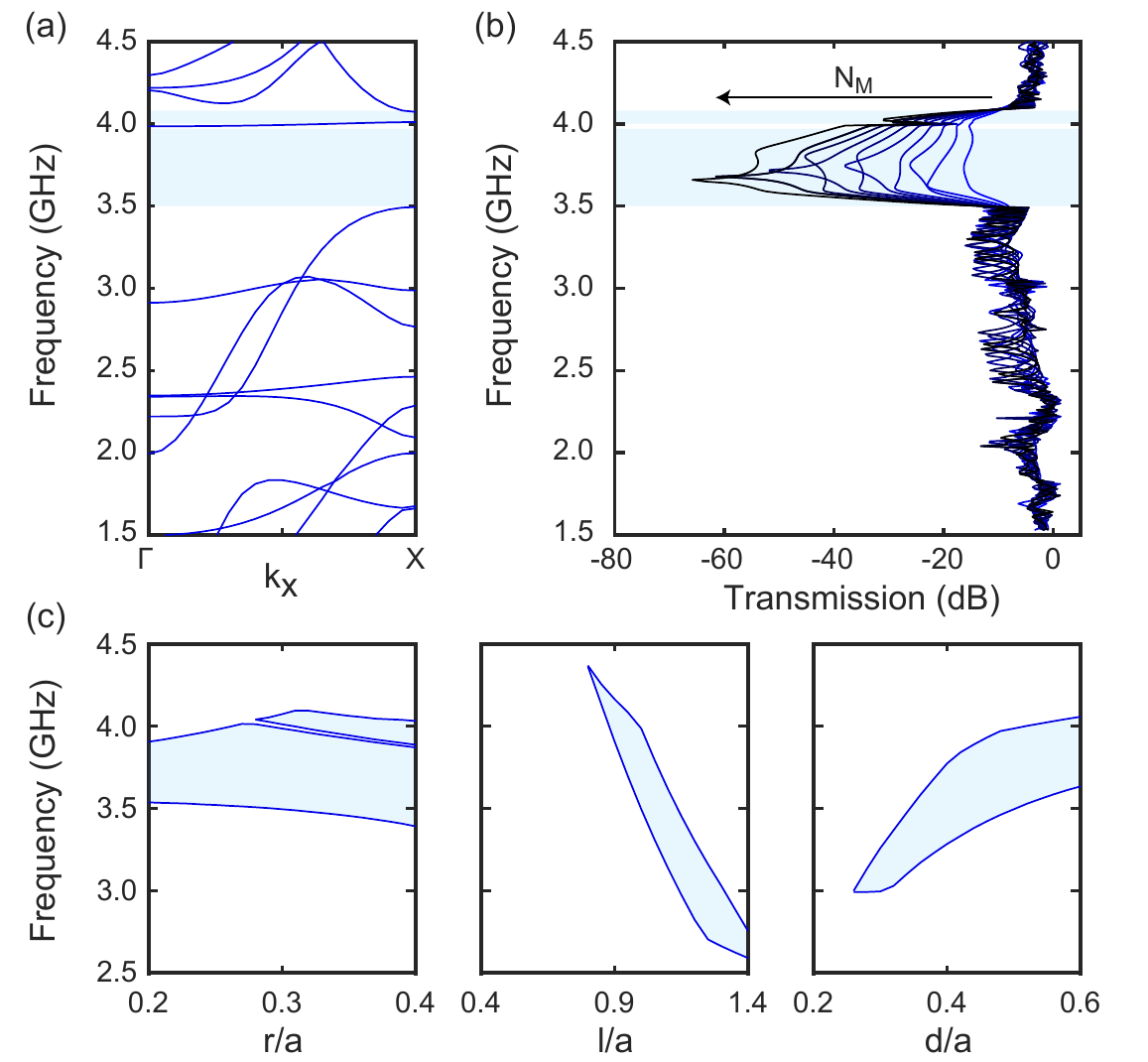}
    \caption{Phononic band calculations of the mirror unit cell. (a) Phononic band diagram of the mirror unit cell for a lattice period $a$=500 nm, $t$=220 nm, $l/a=$0.9, $d/a=$0.4 and $r/a=0.3$. (b) Transmission acoustic spectra for an array of multiple mirror cells ranging from 5 to 15 mirrors. (c) Evolution of the total phononic bandgap for the same aspect ratios shown in Fig. \ref{fig:photobands}(c).}
    \label{fig:phonobands}
\end{figure}

Once we have the appropriate lattice period to confine the optical mode at the target wavelength, we study the mechanical properties of the mirror cell. One of the main advantages of this unit cell profile is that it holds a complete phononic bandgap \cite{OUD14-PRB,MER20-NN}, as can be seen in Fig. \ref{fig:phonobands}(a). Additionally, an analysis in the phononic transmission spectra for an array of multiple mirror cells is shown in Fig. \ref{fig:phonobands}(b). Here, a higher number of mirror cells provides deeper transmission gaps, as expected from the existence of a phononic bandgap. Finally, the evolution of the full phononic bandgap for different aspect ratios is presented in Fig. \ref{fig:phonobands}(c). Again, the main advantage can be seen in the tunability of the system ranging almost all the microwave S-band (between 2 and 4 GHz). More details about the mechanical calculations are provided in the Supporting Information (Fig. S1).

\section{Design of the multimode OMCC}

\begin{figure}
    \centering
    \includegraphics[width=\columnwidth]{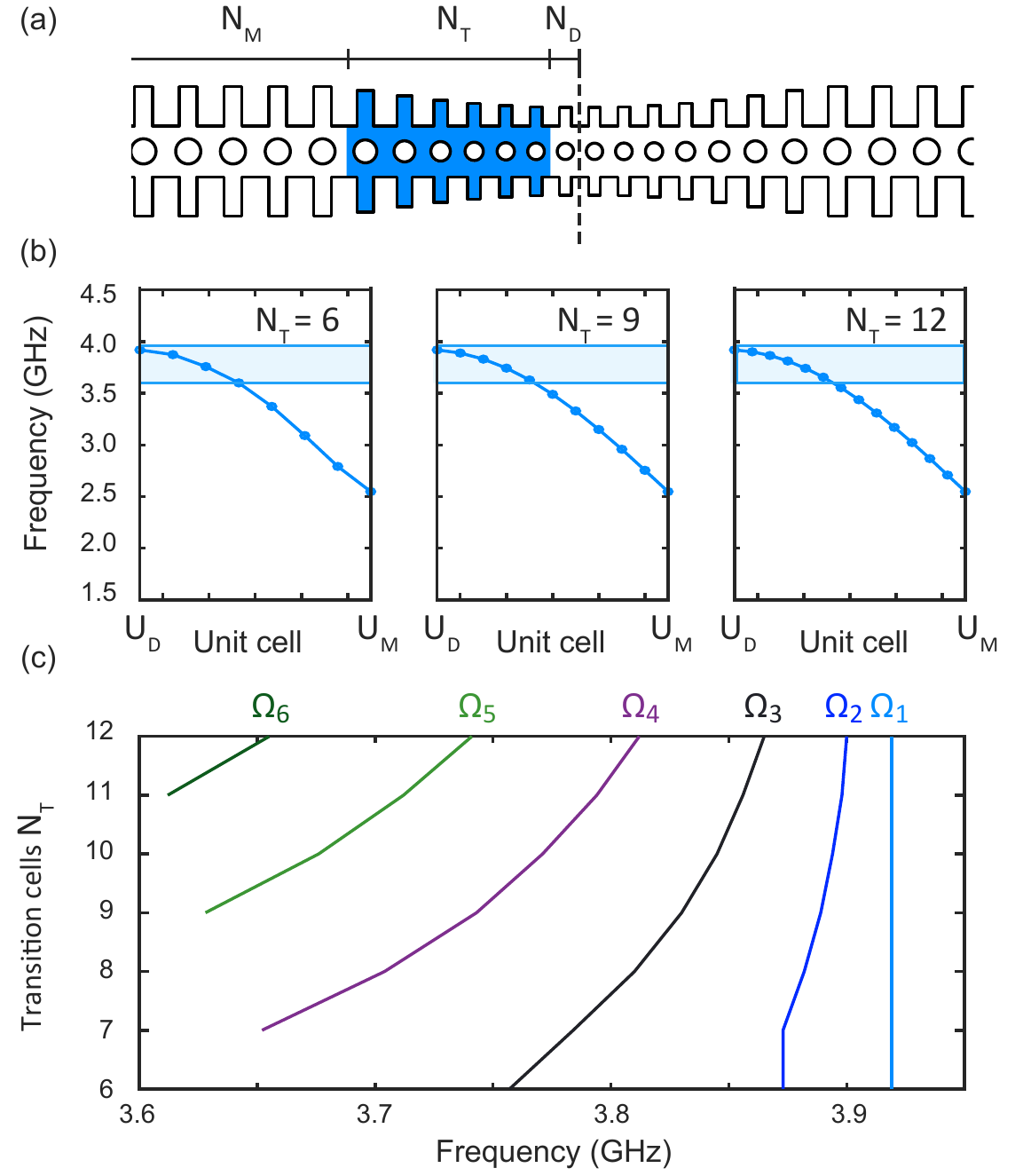}
    \caption{Multimode mechanical design. (a) Sketch of the silicon OMCC used in this work. (b) Evolution of the confined mechanical band at the $\Gamma$ point for different cavity lengths as a function of the analyzed unit cell. (c) Estimation of the mechanical modes lying into the total bandgap for OMCC with different transition cells. The predicted frequency corresponds to the mechanical frequency of the confined band for each defect and transition cells lying into the total bandgap. }
    \label{fig:crystal_design_v1}
\end{figure}

After obtaining the mirror unit cell through the calculations of the photonic and phononic band diagrams, we have tailored the defect and transition unit cells to be able to confine multiple mechanical modes inside the phononic bandgap. Previous designs just focused on a cavity where the mechanical mode was mostly confined into the lateral corrugations of the defect cell \cite{OUD14-PRB,MER20-NN}. However, other mechanical modes, located in other lateral corrugations, could also lie in the bandgap. Figure \ref{fig:crystal_design_v1}(a) shows a top view of the silicon OMCC that we consider in this work, including the labeled number of transition cells that will be under study. As shown in Fig. \ref{fig:phonobands}, this cavity presents a full phononic bandgap for mechanical modes in at frequencies around 3-4 GHz (chosen by design). In order to understand the existence of multiple GHz mechanical modes in a single OMCC and with the final aim of setting a route towards a general method, we obtain the phononic band diagrams of the different unit cells in the transition region between the cavity center and the mirrors (blue region in Fig. \ref{fig:crystal_design_v1}(a)). Depending on the total number of transition cells (N$_{T}$), there will be a certain number of mechanical modes confined in the OMCC with frequencies in the phononic bandgap.

Figure \ref{fig:crystal_design_v1}(b) shows the evolution of the frequency  of the mechanical mode inside the bangap at the $\Gamma$ point for different unit cells that represent the transition from the central defect (U$_{D}$) to the side mirror (U$_{M}$), at each constitutive cavity cell. The panels correspond to OMCCs with 6, 9 and 12 transition cells, respectively. This suggests that, when forming the OMCC, mechanical modes will appear at the calculated frequencies (depicted with dots) inside the total bandgap (drawn with a shaded area) and localized in the unit cells that form the transition between the center of the cavity and the mirrors. 

 
 Remarkably, as the total number of transition cells (N$_{T}$) increases, more mechanical modes appear inside the phononic bandgap, though the increase is not linear. Consequently, an OMCC created with more transition cells should result in more confined mechanical modes. This can be appreciated in Fig. \ref{fig:crystal_design_v1}(c), which shows the evolution of the $\Gamma$ point frequencies of the mechanical modes inside the full phononic bandgap as a function of N$_{T}$. Furthermore, the mechanical frequencies get closer as more transition cells are added, since the physical dimensions between successive cells change more smoothly.
 
\begin{figure}
    \centering
    \includegraphics[width=\columnwidth]{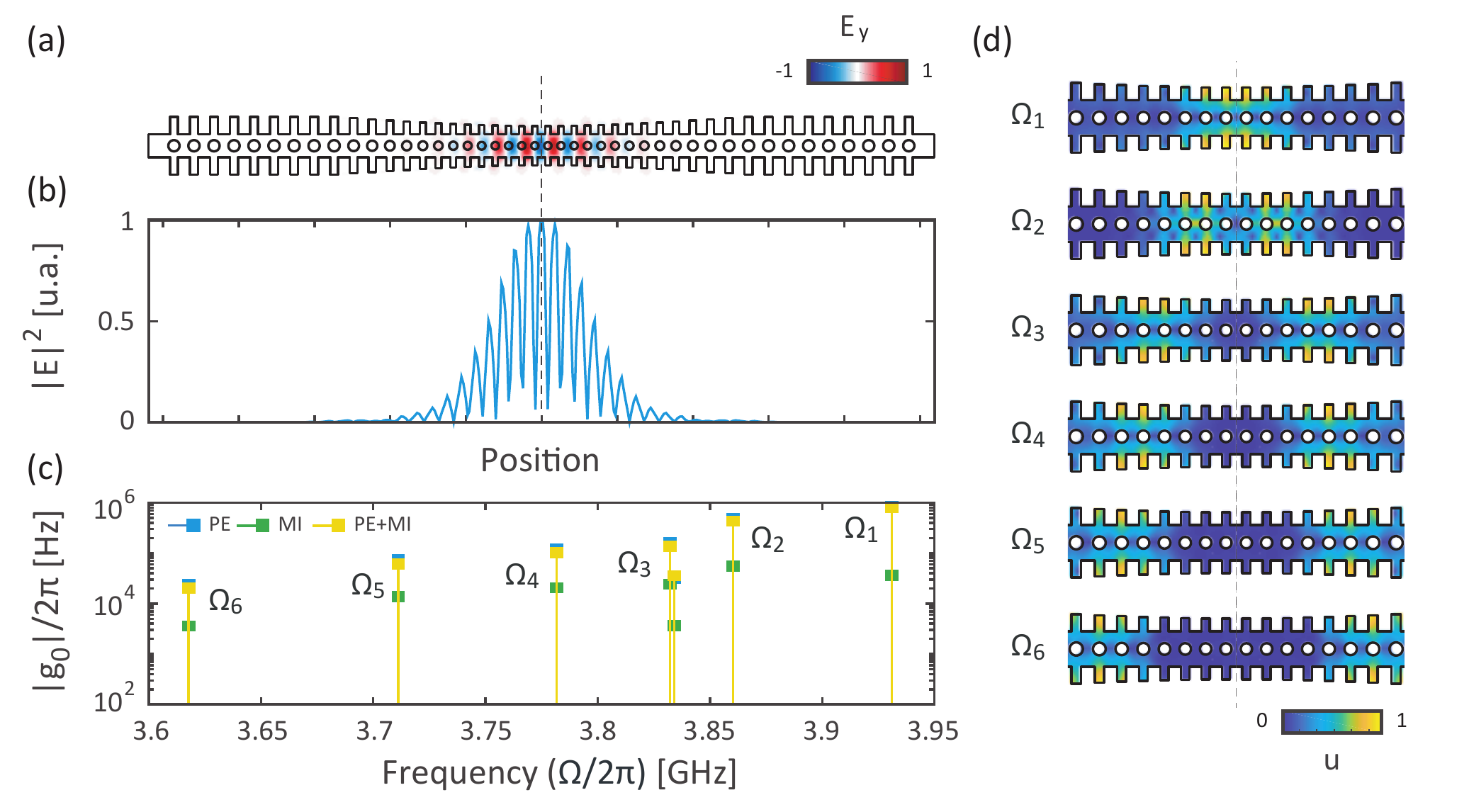}
    \caption{Optical and mechanical modes and OM coupling rates in the OMCC. (a) Electric field profile for an OMCC with 12 transition unit cells at a wavelength of 1570 nm. (b) Normalized total electric field variation along the total length of the cavity. (c) Calculated OM coupling rate contributions for the six fundamental mechanical modes within the phononic bandgap. (d) Displacement field profile for the six mechanical modes giving the largest $g_0$.}
    \label{fig:exp_12cav}
\end{figure}

The next question to address is if all mechanical modes are well coupled with the optical field. To this end we calculate the OM coupling rates $g_0$ of the different mechanical modes of an OMCC with $N_T=12$. Figure \ref{fig:exp_12cav}(a) shows the optical mode of this OMCC. The electric field pattern shows a strong localization in the center of the cavity whilst the intensity exponentially decays as we move toward the mirror regions (see Fig. \ref{fig:exp_12cav}(b)). Once the optical mode is obtained, we separately simulate the fundamental mechanical modes of the OMCC. As it was explained before, not all the mechanical modes display large values of $g_0$. Hence, for all of them we calculate the OM coupling rate $g_0$ with the optical field obtained in \ref{fig:exp_12cav}(a) to find the ones that have the largest $g_0$. Figure \ref{fig:exp_12cav}(c) depicts the six mechanical modes that have the highest OM coupling rates. We consider both the photo-elastic (PE) as well as the moving interface (MI) effects \cite{PEN14-NP} that contribute to the total $g_0$. It is worth noting that the six mechanical modes correspond to the ones predicted in Fig. \ref{fig:crystal_design_v1}(c). Finally, Fig. \ref{fig:exp_12cav}(d) gathers the mechanical profiles of those modes, which are localized at the lateral corrugation of the defect and transition cells. Indeed, the first mode shows localization in the center of the cavity whilst the mechanical localization is displaced toward the mirror regions for higher-order modes. Noticeably, we get relatively large $g_0$ values even when the mechanical modes are localized close to the mirror regions and far from the cavity center.

\begin{figure*}
    \centering
    \includegraphics[width=\textwidth]{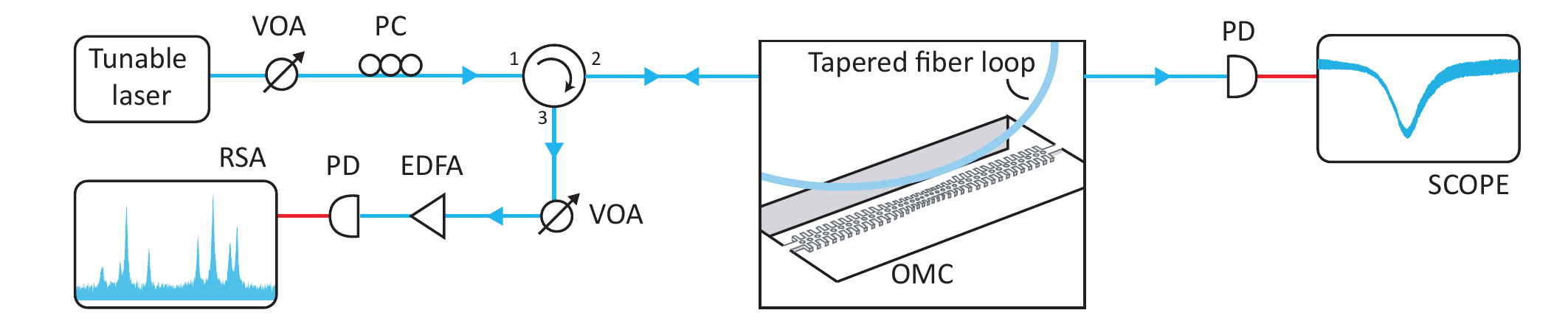}
    \caption{Experimental setup for the optical and mechanical characterization of the OMCCs. VOA: variable optical attenuator, PC: polarization controller, EDFA: erbium doped fiber amplifier, SCOPE: oscilloscope, RSA: radiofrequency spectrum analyzer.}
    \label{fig:exp_setup}
\end{figure*}

\section{Experimental results}

The optical and mechanical experimental characterization of a set of fabricated OMCCs with different transition cells has been performed with the setup sketched in Fig. \ref{fig:exp_setup}. Here, a tunable laser is fed into the system and a variable optical attenuator (VOA) and a controller polarizer are used to set the required input power and polarization in the experimental measurements. Then, the optical signal is sent through an optical circulator. From port 2, the optical signal arrives to a tapered fiber loop which couples to the OMCC via evanescent field coupling. The transmitted signal arrives to a low frequency photodetector connected to the oscilloscope to measure the optical response of the system. On the other side, the reflected signal returns to the optical circulator (port 3) and its optical power is regulated by means of a VOA and an erbium doper fiber amplifier (EDFA). Finally, the output signal (with the mechanical modes transduced on the optical drive) is photodetected with a 12 GHz band photoreceiver and processed with a radiofrequency spectrum analyzer (RSA).

\begin{figure}
    \centering
    \includegraphics[width=\columnwidth]{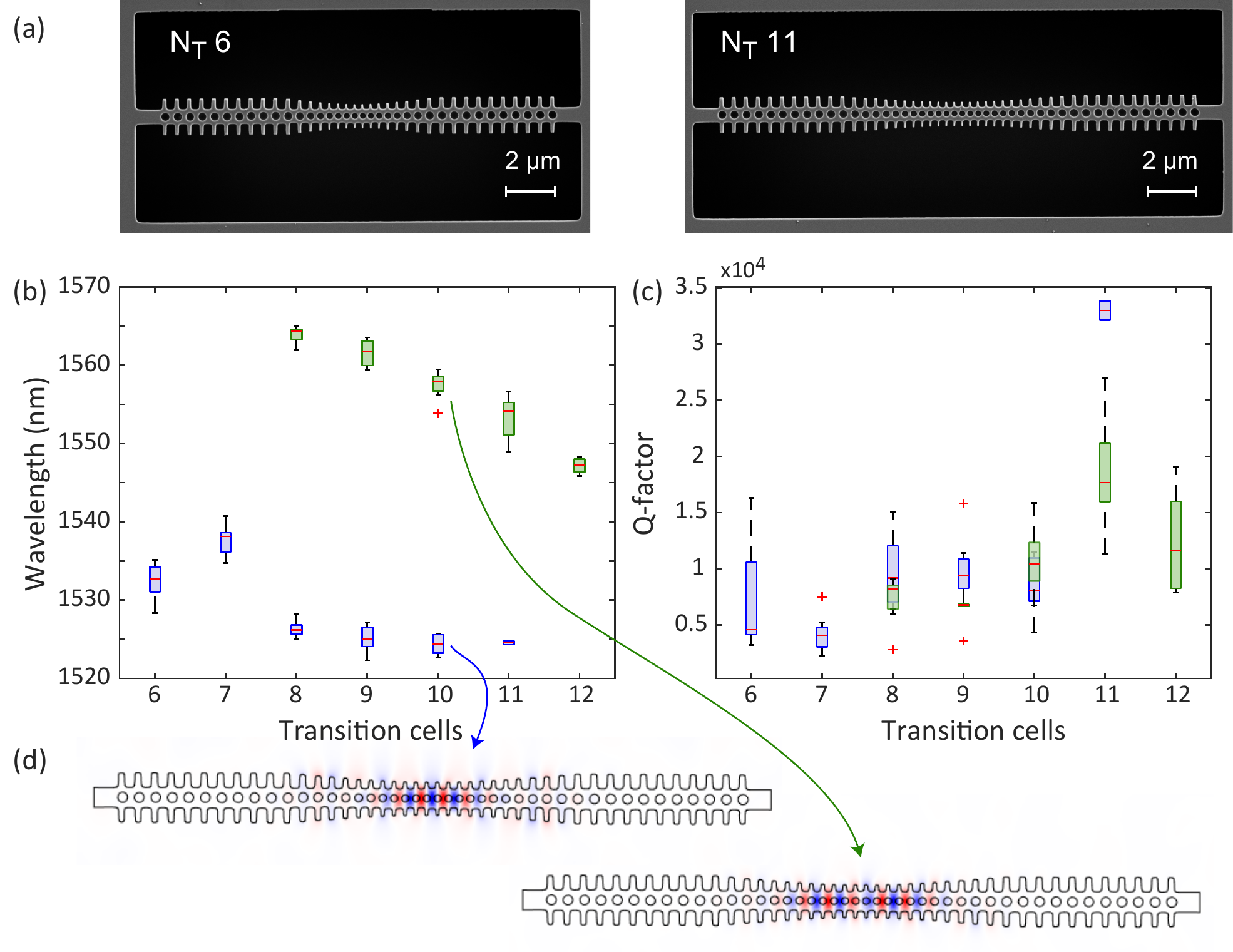}
    \caption{Optical characterization of different OMCCs. (a) Scanning electron microscopy (SEM) images of a fabricated optomechanical crystal cavity with 6 and 11 transition cells. (b) Evolution of the measured wavelength of the two detected optical modes as a function of the number of transition cells. (d) Optical quality factor as a function of the number of transition cells of the cavity. (d) Optical field profiles ($\textit{E}$) for the first and second measured optical modes calculated by retrieving the OMCC pattern from an SEM image.}
    \label{fig:optics}
\end{figure}

Regarding the optical response, the measurements were performed under a low laser input power in order to prevent the appearance of thermo-optic effect, which may result in a bistability asymmetric ”saw-tooth” shaped transmission that gives rise to a shift in the optical resonance \cite{NAV14-AIPA}. For a given design, a set of 12 OMCCs for each number of transition cells was fabricated and characterized. The fabrication process is described elsewhere \cite{MER20-NN}. Figure \ref{fig:optics}(a) shows two scanning electron microscope (SEM) images of fabricated OMCCs having 6 and 11 transition cells OMCCs. In comparison with the designed cavity, a fabrication-induced rounding of the lateral corrugations is clearly appreciated. The optical response was studied for each fabricated cavity as shown in Fig. \ref{fig:optics}. The signal obtained from characterizing each optical mode was a symmetric resonance where a Lorentzian fit was performed to retrieve the optical frequency and quality factor. As shown in Fig. \ref{fig:optics}(b), the cavity supported two optical modes, whose wavelength decreases as more transition cells are added to the cavity. On the other hand, the quality factor increases with $N_T$ for both modes as seen in Fig. \ref{fig:optics}(c), with some exceptions that can be attributed to fabrication irregularities. Figure \ref{fig:optics}(d) represents the electric field profile of both optical modes, which are localized at the defect and transition cells. These profiles were obtained by retrieving the real fabricated pattern from the SEM images. 

\begin{figure}
    \centering
    \includegraphics[width=\columnwidth]{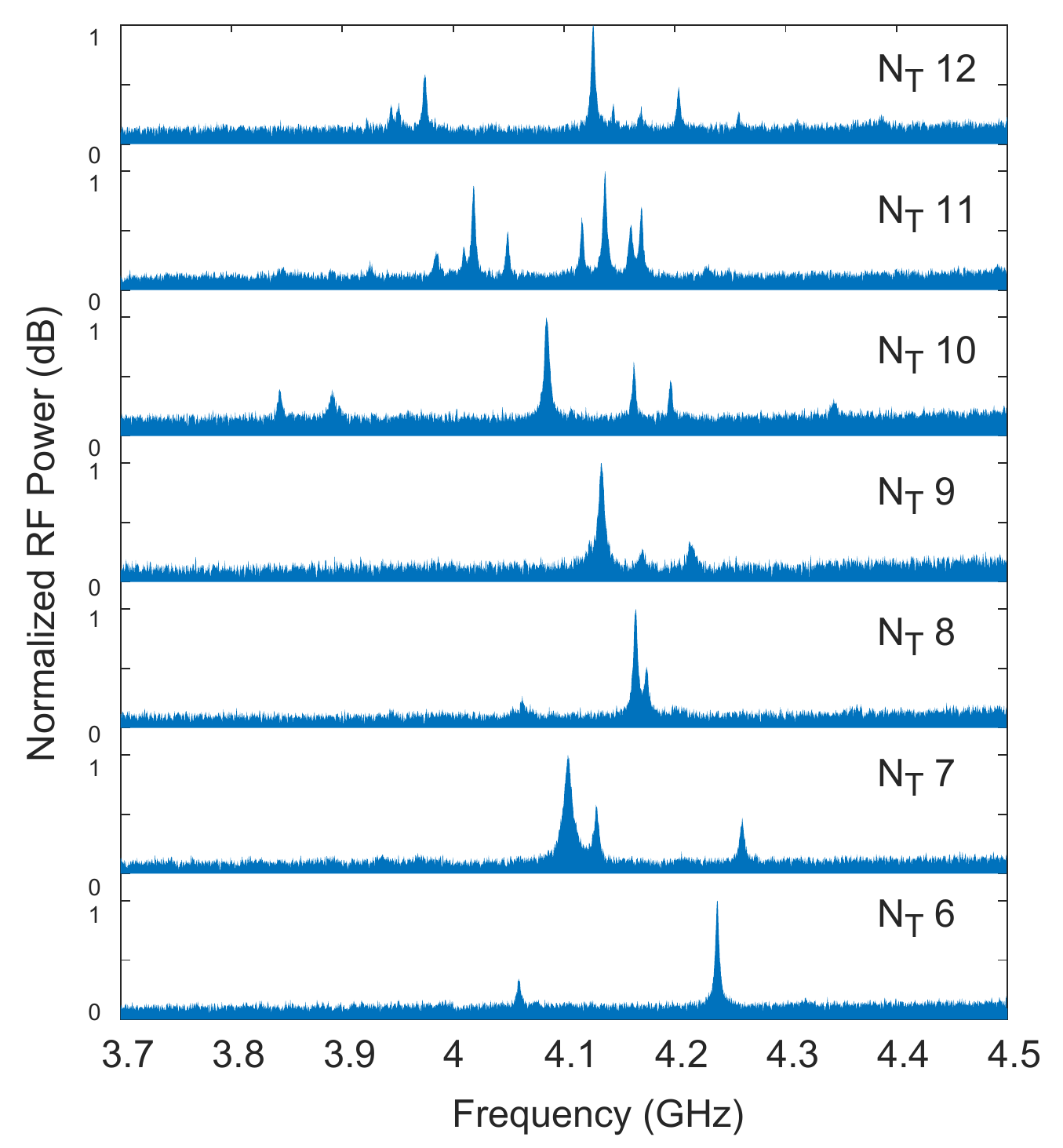}
    \caption{Mechanical response of the OMCCs. Experimental mechanical mode response (represented as a normalized RF power) for different OMCCs as a function of the number of transition cells.}
    \label{fig:mechanics1}
\end{figure}

Concerning the mechanical response, a typical evolution of the measured normalized radiofrequency (RF) spectra showing the mechanical response as a function of the number of transition cells is depicted in Fig. \ref{fig:mechanics1}(a). The increase in the number of mechanical modes with $N_{T}$ predicted in the simulations is also observed in the experiments. Although we did not find all predicted modes, probably because fabrication imperfections (as shown in Fig. \ref{fig:optics}(a)), the measured frequencies are around 4 GHz, which is close to the simulated values and ensures us that the modes come from the lateral corrugations of the defect and transition cells, since those are the only ones that can vibrate at those frequencies in the OMCC. The experimental mechanical modes obtained for other cavities are included in the Supporting Information. 

\begin{figure}
    \centering
    \includegraphics[width=\columnwidth]{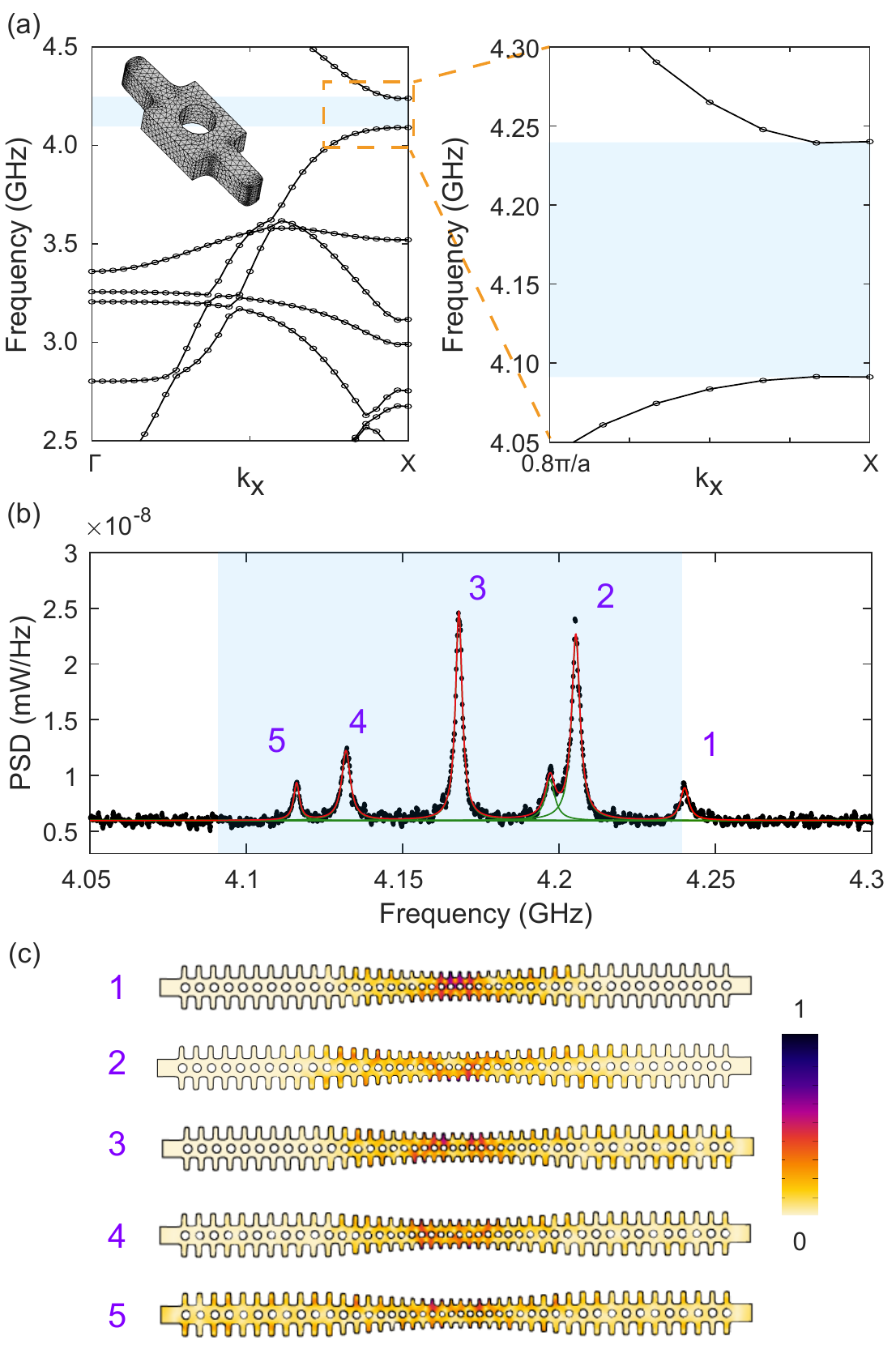}
    \caption{Study of a fabricated OMCC with 10 transition cells. (a) Phononic band diagram of the retrieved SEM unit cells for one of the measured cavities under study. (b) PSD spectrum measurements for a cavity with 10 transition cells. (c) Retrieved SEM mechanical mode profiles from the fabricated structure.}
    \label{fig:mechanics_sem}
\end{figure}

A detailed analysis of the characterization of an OMCC with 10 transition cells is presented in Fig. \ref{fig:mechanics_sem}. First, an analysis of the phononic band diagram for the fabricated structure was performed. As one of the main objectives here is to confine the mechanical modes into a total bandgap, we retrieved from the SEM images the real profile for the different unit cells of the mirror cavity and calculated the expected band diagram. The result can be seen in Fig. \ref{fig:mechanics_sem}(a), where an inset of a mirror unit cell of the measured OMCC is presented. As expected, the cavity shows a full phononic bandgap around 4 GHz and the measured mechanical modes lies into it, as shown in Fig. \ref{fig:mechanics_sem}(b). Here, it can be seen that, even after fabrication imperfections, it is quite in accordance with the mechanical modes predicted with the band diagram shown in Fig. \ref{fig:crystal_design_v1}(c). However, it must be noted that double peaks as the one corresponding to peak 2 can appear as a result in difference between the fabricated corrugations in cells of the same nominal dimensions. To ascertain the mechanical mode profiles of each measured peak, we retrieved and simulated the actual fabricated OMCC profile from its SEM image. Figure \ref{fig:mechanics_sem}(c) shows the mechanical mode profiles for the retrieved profile showing that, despite the mechanical motion is not totally confined in a single corrugation, as in the nominal cavity in Fig. \ref{fig:exp_12cav}, the position of mechanical displacement moves toward the extremes of the cavity as the frequency decreased, as expected from the numerical modelling.


\section{Conclusions}

In this work, we have proposed and demonstrated a method to engineer multiple mechanical modes with GHz frequencies within a full phononic bandgap in a silicon OMCC. The OMCC is formed by drilling circular holes and adding lateral corrugations to a released silicon nanobeam. By increasing the number of cells in the adiabatic transition between the cavity center and the lateral mirrors, more and more mechanical modes appear in the cavity. All the mechanical modes show reasonably large values of $g_0/2\pi$ up to 600 kHz which enable efficient transduction into a driving optical signal. Multiple applications can be envisaged, including multimode phonon lasers \cite{MER21-PRL}, frequency up- and down-conversion of multiple wireless signals \cite{MER21-LPR} or building chiral nano-optomechanical networks \cite{PINO22-NAT}.

\appendix

\section{Phononic band diagram and transmission simulations}

\begin{figure}[h!]
    \centering
    \includegraphics[width=\columnwidth]{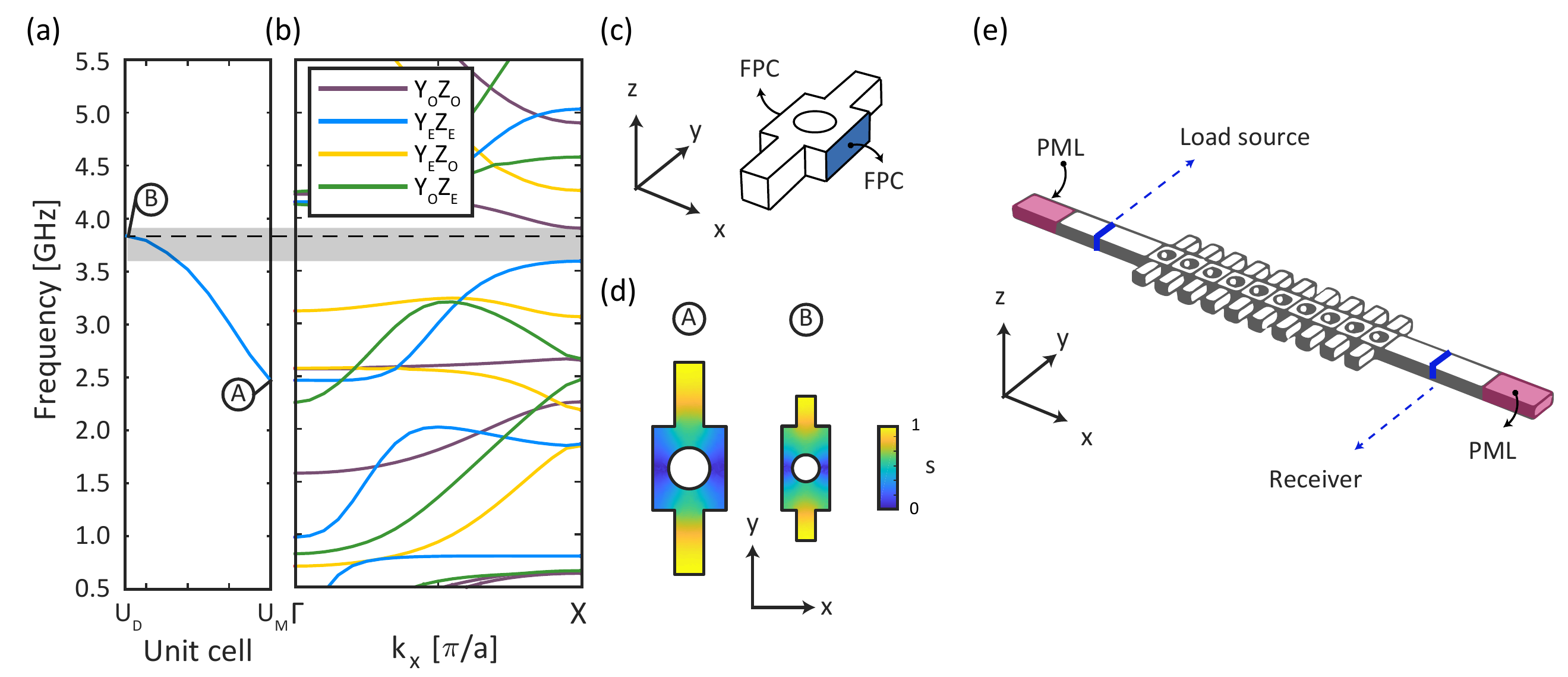}
    \caption{(a) Phononic band diagram evolution of the confined mechanical mode band from the defect (UM) to the mirror (UD) unit cell. (b) Phononic bands for mechanical modes with (odd, odd) in purple, (even, even) in blue, (even, odd) in yellow and (odd, even) in green symmetries respect to the Y and Z axis planes. The gray-shaded area denotes the total phononic bandgap and the confined mechanical mode in the defect zone has been depicted with a dashed line. (c) Mechanical unit cell showing the boundary Floquet Periodic Condition (FPC) of the simulation. (d) Mechanical field profile of (UM) in A and (UD) in B. (e) Tranmission phononic simulations scheme. The PMLs are shadowed in pink and the source and receiver areas are depicted in blue.
}
    \label{fig:phonosimus}
\end{figure}

Simulations of phononic bands were performed with COMSOL Multiphysics \cite{comsol}, as in previous works \cite{MER20-NN}. The evolution of the frequency at at the $\Gamma$ symmetry point for the studied unit cell changing the parameters from the central defect to the mirror region is presented in Fig. S\ref{fig:phonosimus}(a). Figure S\ref{fig:phonosimus}(b) shows in more detail all the involved bands $-$ including different symmetries $-$ for the mirror unit cell. The most important feature to emphasize is the existence of a complete phononic bandgap where, by proper design, the frequencies of the engineered confined mechanical modes have to be placed. As noted in the main text, this should reduce the phonon leakage of the final structure as it prevents that the confined mechanical mode couples with modes of different symmetries existing in the mirror regions. In these simulations, we set Floquet Periodic Conditions (FPC) at the lateral boundaries of the structure, as shown in Fig. S\ref{fig:phonosimus}(c), and the remaining boundaries were kept as free. The dimension values of the mirror and the unit cell are the same as the ones used in the main text. Regarding the estimation of the number of expected mechanical modes to be confined in cavities with different transition cells, the analysis was similar to the one performed in S\ref{fig:phonosimus}(a). Here, the point is that, as the number of transition cells used to build up the cavity increases, there will be more unit cells with dimensions close to those of the center region with frequencies within the bandgap. Because of that, as presented in Fig. 3(b) in the main text, we can estimate the total number of mechanical modes just by analyzing the eigenfrequencies and eigenvalues of each unit cell. It must be noted, however, that the cavity may also support less localized mechanical modes, as presented in Fig. 3(d), but, as shown in the same figure, the estimation of the number of confined mechanical modes is still correct. 

Regarding the phononic transmission simulations, an scheme of the boundary conditions of the system is presented in Fig. S\ref{fig:phonosimus}(e). Here, we simulated the transmission coefficient of a signal generated at the load source (set as a boundary source in this case) and received at the receiver area. The transmission coefficient was calculated as the ratio total displacement ($\textbf{u}$) integrated in the source and the receiver as $\int_{Receiver}(\textbf{u})/\int_{Source}(\textbf{u})$. At the extreme boundaries of the system we set perfectly matched layers and the rest of the boundaries were kept as free. This simulation was also performed with COMSOL Multiphysics.

\section{Measurement of the transduced mechanical modes}

In the experiments, we were able to test OMCCs with different values of $N_{T}$, all with the same nominal parameters according to our simulations. Figure S\ref{fig:mechanics1} shows the measured RF spectra for different values of $N_{T}$ of cavities with the same nominal parameters as those reported in the main text but fabricated with a different e-beam dose. The appearance of multiple mechanical modes within the phononic bandgap is also evident here.  

We also measured the mechanical Q factor of the transduced mechanical modes. The results are shown in Fig. S\ref{fig:mechanics_q}, which represents the mechanical Q factor of the different mechanical modes for each value of $N_{T}$. The mechanical quality factor was evaluated from the ratio between the center frequency of the peak and mechanical linewidth. The presented values were obtained through a lorentzian fit of each peak, as can be seen in the green fits in Fig. S\ref{fig:mechanics1} for different cavity lengths. In all these panels we can also presented in red the total fit envelope of the system. It can be seen that the average value is around 1000, as expected in this kind of cavity when operated at room temperature. 

\begin{figure}[h!]
    \centering
    \includegraphics[width=\columnwidth]{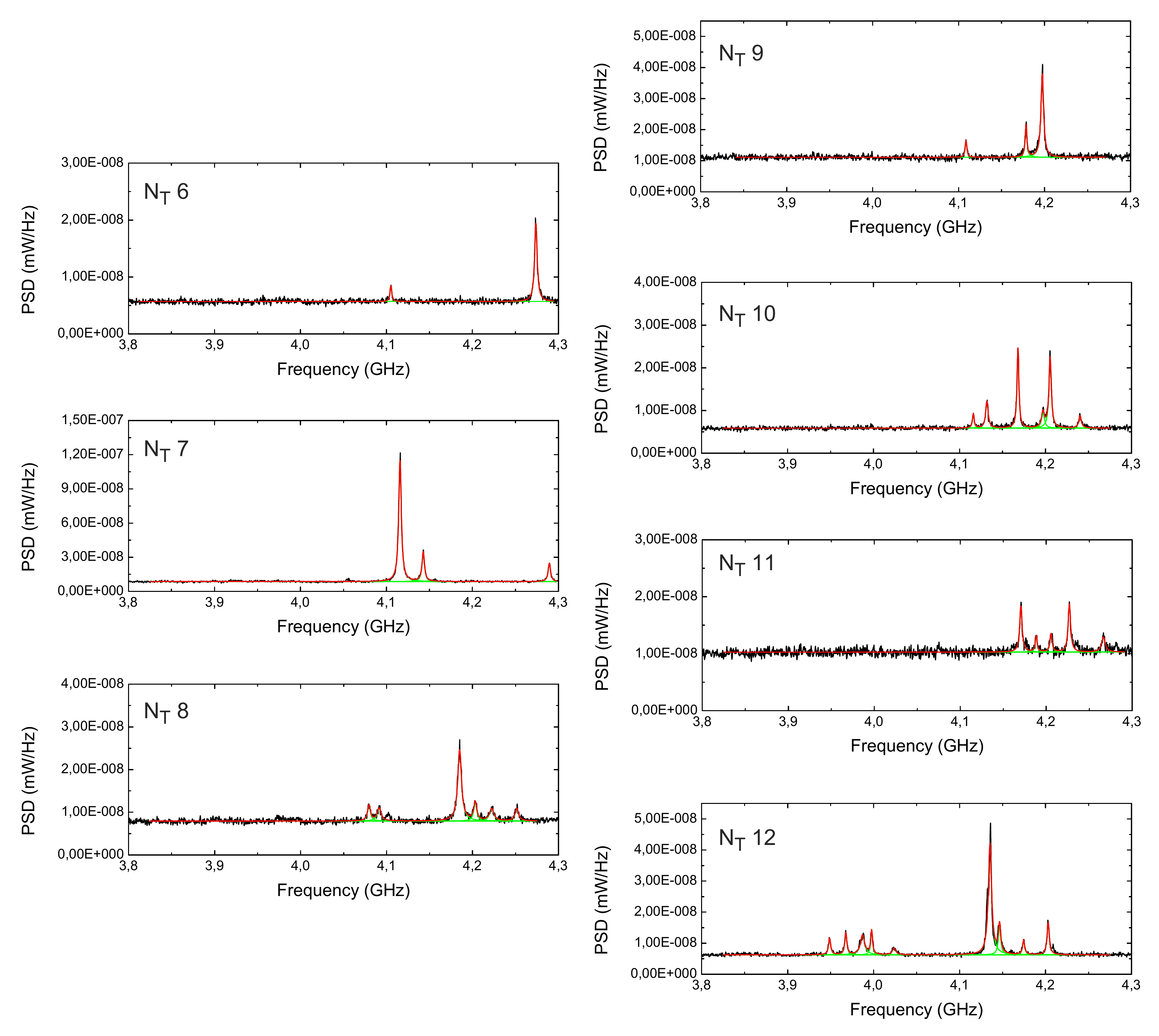}
    \caption{Experimental mechanical mode response for different OMCCs as a function of the number of transition cells. In green are presented the lorentizian peaks of each individual mechanical mode and in red the resulting envelope fit.}
    \label{fig:mechanics1}
\end{figure}

\begin{figure}[h!]
    \centering
    \includegraphics[width=\columnwidth]{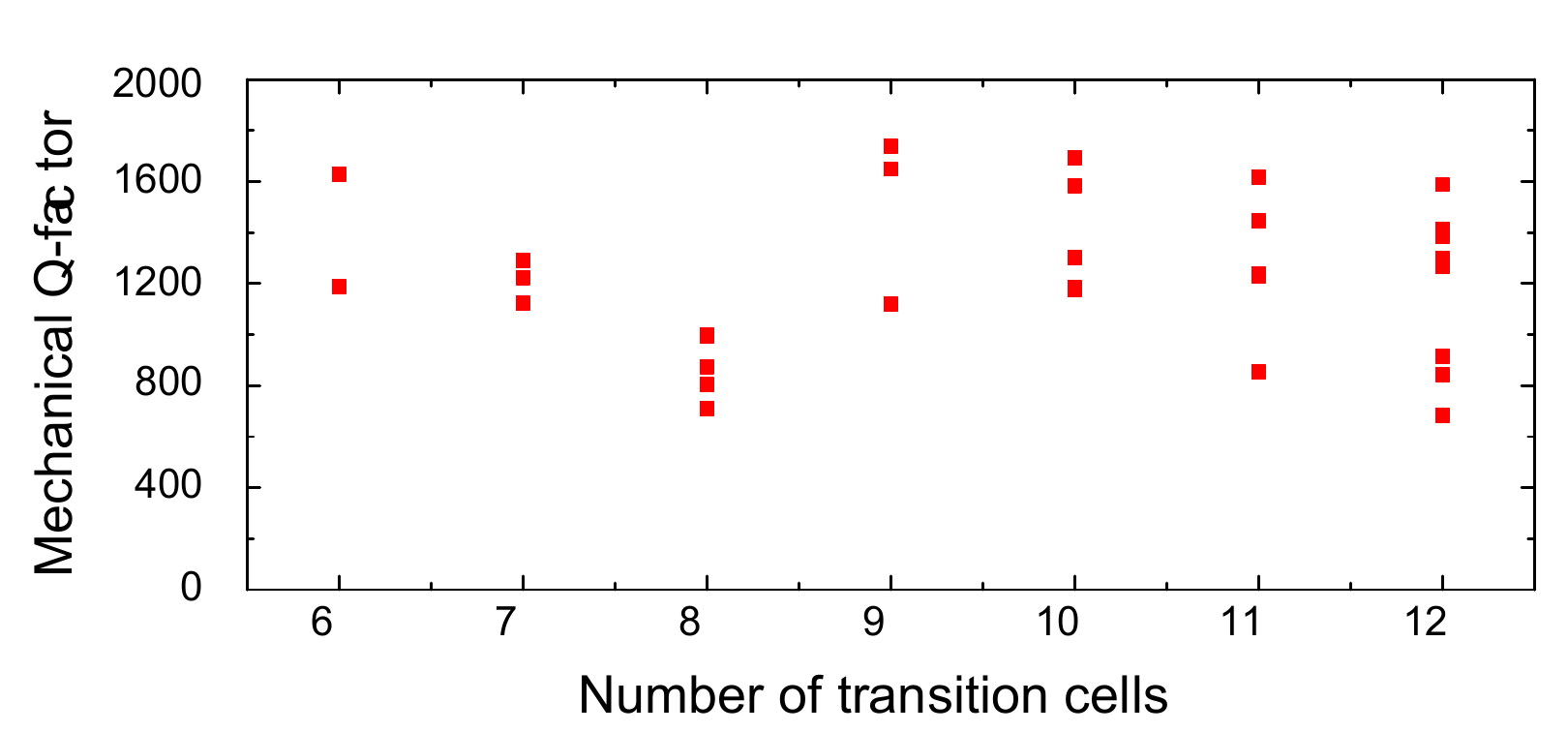}
    \caption{Retrieved experimental mechanical quality factors from the lorentzian fits of the transduced peaks shown in Fig. S\ref{fig:mechanics1}.}
    \label{fig:mechanics_q}
\end{figure}

\begin{figure}[h!]
    \centering
    \includegraphics[width=\columnwidth]{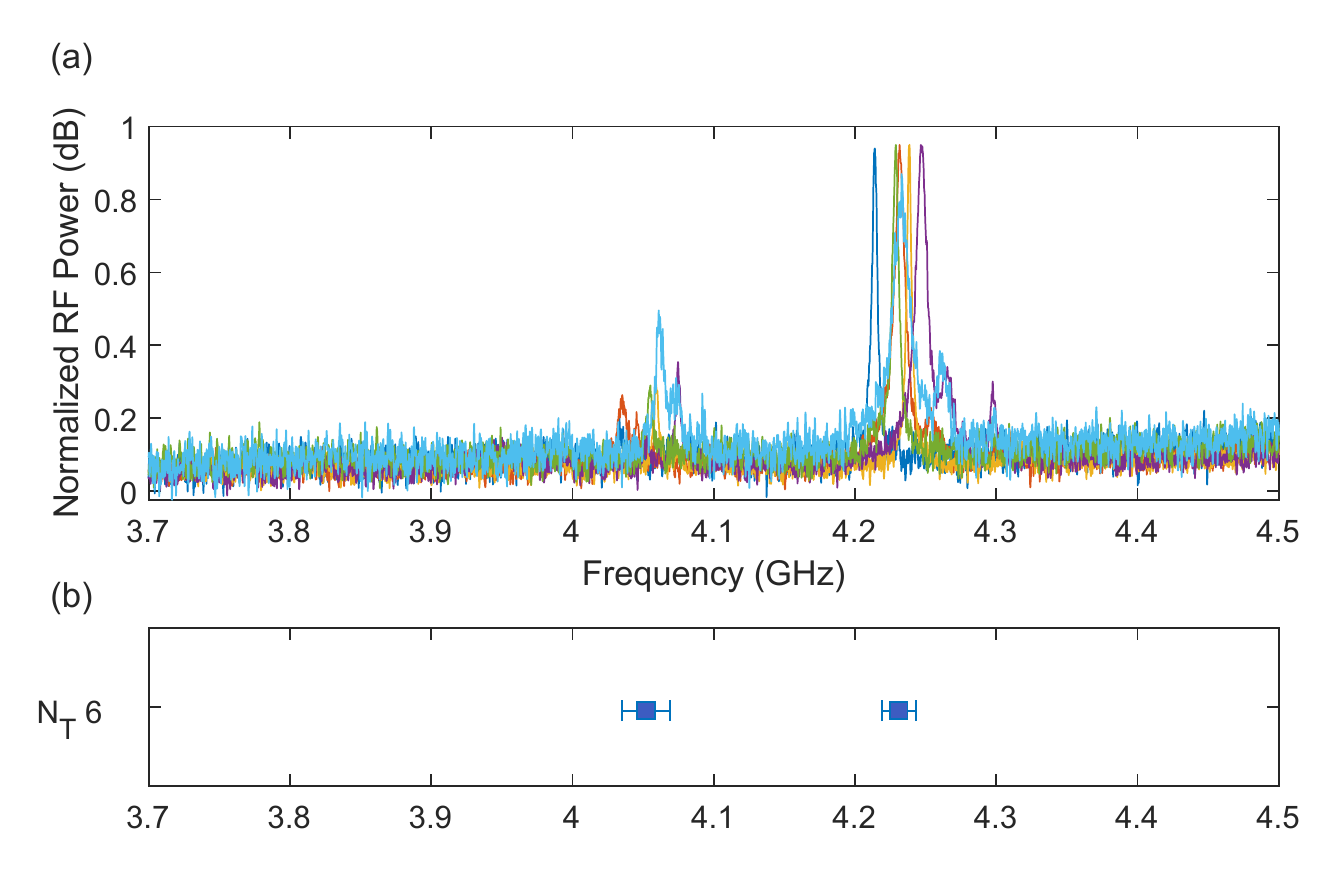}
    \caption{(a) Normalized RF power spectra of multiple measurements of cavities with 6 transition cells nominally equal. (b) Mean value and standard deviation of the two main peaks for a cavity with 6 transition cells for a statistic sample of 12 cavities. }
    \label{fig:mechanics_tc6}
\end{figure}

We also compared the response of OMCCs with the same $N_{T}$ and identical nominal values, fabricated with the same e-beam does. The results for different fabricated OMCCs having $N_{T}=6$ are depicted in Fig. S\ref{fig:mechanics_tc6}(a) for a set of 6 measurements, for the sake of clarity. It can be observed that the spectral dispersion is low, as seen in Fig. S\ref{fig:mechanics_tc6}(b) that shows the mean value and the standard deviation for a set of 12 cavities.

\begin{acknowledgments}
The authors acknowledge funding from the Spanish State Research Agency (PID2021-124618NB-C21 and PID2021-124618NB-C22); Generalitat Valenciana (BEST/2020/178, PROMETEO/2019/123, IDIFEDER/2020/041 and IDIFEDER/2021/061). L.M. thanks financial support from the Next generation EU program, Ministerio de Universidades (Gobierno de España).
\end{acknowledgments}

\section*{Data Availability Statement}

The data that support the findings of this study are available from the corresponding author upon reasonable request.

\bibliography{biblio}


\end{document}